\documentclass{article}

\begin{document}

\title{Advancements in solar neutrino physics}

\author{
VITO ANTONELLI, LINO MIRAMONTI\\
Physics Department of Milano University and INFN\\ 
via Celoria 16, Milano, Italia I-20133.\\\
}

\maketitle

\begin{abstract}
We review the results of solar neutrino physics, with particular attention to the data obtained and the analyses performed in the last decades, which were determinant to solve the Solar Neutrino Problem, proving that neutrinos are massive and oscillating particles and contributing to refine the Solar Models. We discuss also the perspectives of the presently running experiments in this sector and  of the ones planned for the near future and the impact they can have on elementary particle physics and astrophysics.
\end{abstract}

%\keywords{Solar neutrinos.}

\section{Introduction } 

Solar neutrinos study has given, since ever, a fundamental contribution to 
%the development of 
elementary particle physics and astrophysics, 
%contributing to 
creating a link between these two disciplines.  The original idea of Bahcall and Davis to use neutrinos as  probes to investigate the Sun's properties opened half a century ago an unexpected scenario, giving rise to the long standing \lq\lq Solar Neutrino Problem\rq\rq, which had a great impact on our knowledge of particle physics. The answer to this puzzle, found around the turn of the millennium combining the data obtained by the radiochemical experiments, by SuperKamiokande and more recently by SNO and the reactor experiment KamLAND, proved in a crystal clear way that neutrinos are massive and oscillating particles, showing the need to go beyond the original version of the Standard Model and offering a significant test for every theory \lq \lq beyond the Standard Model\rq\rq. 

In the last years a change of paradigm took place in neutrino physics and the attention is now mainly focused on appearance experiments and on the study of artificial sources (short and long baseline reactor and accelerator experiments and superbeams), 
which are expected to find at least partial answers to the many open questions of neutrino physics. However, solar neutrino physics 
still can contribute significantly to the search of some of these answers. In particular it can help to improve and complete the knowledge of mass and mixing pattern, to study, for the first time with real time experiments, the low energy part of solar neutrino spectrum and look for possible anomalies in the oscillation pattern. The first step in this direction has been moved by the Borexino 
collaboration and further contributions are expected by this experiment and also by SNO+ and other experiments planned for the near and far future.
 
The huge quantity of data accumulated and the accuracy reached by many of the different solar neutrino experiments brought back to new life the original purpose of solar neutrino studies, that is the analysis of the mechanisms regulating the Sun's shining and evolution. In particular it could be important to contribute to solve the so called solar metallicity problem, even if this will be a hard task that requires a simultaneous study not only of the pp-chain, but also of the almost unexplored {\it CNO} bi-cycle.  

In this paper, after a brief review of the milestones of solar neutrino history, 
%(from the pioneristic Homestake experiment to the present results),
with particular attention to the results of the last 10-15 years, we discuss the status of solar neutrino mixing and oscillation in the three flavor scenario and comment on possible future perspectives of this sector.  

\section{From the rise of the Solar Neutrino Problem to its solution}
The history of solar neutrino studies started in the latest 1960 in  South  Dakota, when the  first  experiment  to  detect solar  neutrinos  was built  in  the Homestake  gold  mine~\cite{Davis1968cp} \cite{Davis1968cp2} \cite{Davis1968cp3} \cite{Davis1968cp4} .  It operated continuously from 1970 until 1994, with a  large detector consisting  in a tank filled with  615 metric tons of  liquid perchloroethylene ${\rm C_{2}Cl_{4}}$, suitable to detect $^{7}Be$  and $^{8}B$, and a small signal from  the {\it CNO} and {\it pep} solar neutrinos, via the reaction $\nu_{e} + {\rm ^{37}Cl} \rightarrow {\rm ^{37}Ar} + e^{-} \,$ with an energy threshold of 
$E_{th} = 814 \: \rm{keV}$. 
%Once a month, after bubbling helium through the tank, the $^{37}$Ar atoms (the radioactive $^{37}Ar$ isotopes  decay by electron capture with a 
%$\tau_{1/2}$ of about 35 days into ${\rm ^{37}Cl^{*}}$) were extracted and counted. 
%The number of atoms created was about $5$ atoms of $^{37}$Ar  per  month. 
The number  of  detected neutrinos  was only about 1/3  of the quantity  expected from the  Solar Standard  Model (SSM) and this large  discrepancy 
%between the neutrinos detected by Homestake and one expected from the SSM i
is the essence  of the Solar  Neutrino Problem (SNP).
In the following years SSM passed all tests that helioseismology (the science that studies the interior of the Sun by looking at its vibration modes) started to  provide and this indicated only two possible solutions to the SNP: Homestake could be wrong,  (i.e. the detector was inefficient) or  something happened to the neutrinos while travelling from the core of the Sun to Earth where they were detected.

In 1982-83 Kamiokande a large  water \v{C}erenkov detector (of 3048 metric tons) was built in Japan~\cite{Kamiokande89}\,. It  was the first example of real time solar neutrino detectors, which study the  \v{C}erenkov light  produced  by the  electrons scattered by an impinging neutrino via the reaction $\nu_{x} + e^{-} \rightarrow \nu_{x} + e^{-} \, $ and by observing the direction of the recoiled neutrino can reconstruct also the incoming neutrino direction.  The energy threshold of  the reaction is $E_{th} = 7.5\; {\rm MeV}$ and, therefore, only $^{8}$B and $hep$  neutrinos were detected.  At the beginning of the '90s  a larger version  of the detector  was built, Super-Kamiokande, where the active  mass was more than 50000 metric tons of pure water, and  the energy  threshold was  lowered to~\cite{SK99}\,$E_{th} =  5.5\; {\rm  MeV}$. The  number of detected  neutrinos was about  1/2 lower  than  the expected ones,  aggravating the  SNP.  
The existence of a Solar Neutrino Problem was confirmed also by two radiochemical experiments, studying solar neutrinos via the reaction $ \nu_{e} + {\rm ^{71}Ga} \rightarrow {\rm ^{71}Ge} + e^{-} \, $ which, thanks to a lower energy threshold ($E_{th} = 233\; {\rm keV}$), made possible the detection also of ${\it  pp}$ neutrinos. They are  the lowest energy and less model-dependent component of neutrino spectra and hence the more robust  to test the hypothesis that fusion of hydrogen powers the Sun.  Both the soviet-american  experiment SAGE, 
%located in the  Baksan underground laboratory and 
using more than 50 metric tons of metallic gallium~\cite{Abd99}\,, and the Gallex experiment  at the Gran Sasso underground laboratory, employing 30 metric tons  of natural gallium~\cite{Hamp99,Alt05}\,, measured a neutrino signal smaller ( $\approx\;60\%$) than predicted  by the SSM. For an historical introduction see for instance~\cite{14-Miramonti:2002wz}\,.

The real breakthrough in solar neutrino physics was  the advent of the Sudbury Neutrino  Observatory - SNO - experiment. It  was able to measure simultaneously,  by means of a deuterium  \v{C}erenkov detector, three different interaction channels. The neutral current $\nu_X + d  \to \nu_X + p^+ +  n$, receiving contributions from  all active flavors, the elastic scattering $\nu_X + e^- \to \, \nu_X + e^- $ and the charged current $\nu_e +  d \to e^-  + p^+  + p^+$, that  is sensitive  only to electronic neutrinos.   
Thank to that it was possible to prove clearly and  in a direct  way that the  measured  total neutrino  flux  was  in very  good agreement with the SSM predictions, but only a fraction of these neutrinos had conserved their flavor during their travel from the core of the Sun to the Earth.
The first  SNO data~\cite{SNOES}\,, published  in  2001,  confirmed the  results  obtained  by the other solar neutrino    experiments, providing a significant  evidence of  the presence of a  non-electronic active neutrino  component in the solar flux. This result indicated, for the first time, the Large Mixing Angle (LMA)  as the preferred solution  of the SNP, even if different alternative possibilities (LOW-solution, for instance) were still surviving.  
The SNO Collaboration measured also the neutral current channel, using  different techniques. The data  collected with these techniques 
%called by the Collaboration phases  
are usually   reported   as  SNO I~\cite{SNOI}\,,  SNO   II~\cite{SNOII}\,,  
%in which they  added 
with the addition of salt  to  improve  the efficiency of  neutral current detection,  and SNO III~\cite{SNOIII}\,, with the use of helium chamber proportional counters.

A fundamental contribution to the solution of the SNP came also from KamLAND. 
All the previous reactor experiments, using neutrino energy beams of the order of the MeV with a baseline of the  order of 1 km,  could test only values of  $\Delta m^2$ above $10^{-3}$  $\rm{eV}^2$.   
The KamLAND  experiment, with an  average baseline of about 180 km, was, instead, ideal to probe the LMA region, which corresponds to values  of $\Delta  \rm{m}^2$ of the  order~\cite{SK_KLpotentialities, mioKL}\,$10^{-5}-10^{-4}  \, \rm{eV}^2$.  The deficit  of events   observed by the KamLAND Collaboration was  inconsistent with  the expected rate  in absence  of oscillation at the $99.95 \%$ confidence level.x
The  LMA solution was the  only oscillation solution compatible with KamLAND results and CPT invariance.
Furthermore KamLAND data also restricted  the allowed LMA region; the preferred values for $\Delta m^2_{12}$ and $\theta_{12}$ are slightly  higher than the ones  corresponding to the best  fit solution of the solar  neutrino experiments,  but this small difference is explained taking into  account the experimental uncertainties. 

\section{Last decade solar neutrino measurements and KamLAND results}\label{last-decade}

In the second phase of SNO, which began in July  2001,  the addition of 2000~kg of NaCl  to the 1000 metric tons of $D_2 O$ increased  by a factor of 3 the efficiency with respect to the pure $D_2 O$ phase in the  detection of neutrons produced in  the neutral current (NC) disintegration of deuterons by solar  neutrinos and it enhanced the  energy of the  $\gamma$-ray  coming  from  neutron  capture. In this way it was possible (also thanks to the different event isotropy of the multiple $\gamma$ ray emission produced by the neutron capture on $^{35}Cl$ with respect to the single electron \v{C}erenkov light emitted  in charged current interactions) to separate the neutral from the  charged current events without any  additional assumption on the neutrino  energy spectrum. The results, reported in~\cite{SNOII_2005}\,, included  the full  data  collected in 391 live days of  the salt phase and were analyzed  in terms of the  CC spectra, with a threshold starting from $5.5 \, \rm{MeV}$ kinetic  energy  and NC  and  ES integrated  fluxes  separately  for day  and night. They confirmed the agreement with the SSM for what concerns the total $^8B$ spectrum, 
favored even larger values of the solar mixing angle and gave no indications of day night asymmetries in NC rate (which would have been a signal of oscillations to sterile neutrinos or non standard interactions).  
%The second phase of SNO began in June  of 2001 (SNO II) in which 2000~kg of NaCl  were added to the 1000 metric tons of $D_2 %O$.  The addition of salt increased  the efficiency by a factor of 3 with respect to the pure $D_2 O$ phase in the  detection of %neutrons produced in  the neutral current (NC) disintegration of deuterons by solar  neutrinos and, by enhancing the  energy of %the  $\gamma$-ray  coming  from  neutron  capture. Thanks to that it was possible (also taking advantage from the different event %isotropy of the multiple $\gamma$ ray emission from neutron capture on $^{35}Cl$ with respect to the \v{C}erenkov light emitted by %the charged current single electron) to separate the neutral from the  charged current events without any  additional assumption % on the neutrino  energy spectrum.  
%The salt  phase results was reported in~\cite{SNOII_2005}, which included  the full  data,  collected in 391 live days, of  the salt %phase, analyzed  in terms of the  CC spectra, with a threshold starting from $5.5 \, \rm{MeV}$ kinetic  energy  and NC  and  ES i%ntegrated  fluxes  separately  for day  and night. 

%The third  SNO phase (SNO III) started in November  2004. The neutral  current signal  neutrons  were detected  with an  array of  % $^3  He$ proportional  counters  deployed  in  the  $D_2  O$ and  looking  at  the  gas ionization  induced by neutron capture  on % $^3$He in order to reduce the fluxes correlation and to improve the accuracy of the determination of the mixing angle.
During the SNO III phase,  started in November  2004,  the neutral  current signal  neutrons  were detected  with an  array of  $^3  He$ proportional  counters  deployed  in  the  $D_2  O$ and  looking  at  the  gas ionization induced by neutron capture  on $^3$He, in order to reduce the fluxes correlation and to improve the accuracy of the determination of the mixing angle. The measured total 
$^8B$ neutrino flux was in substantial agreement with previous measurements and SSM and the ratio of the flux measured with CC and NC turned out to be $\Phi_{CC}/\Phi_{NC} = 0.301\pm0.033$. 

A joint reanalysis of SNO  I and SNO  II data, known as  LETA (Low Energy   Threshold   Analysis)~\cite{LETA}\,,  with   improved calibration  and analysis  techniques, was performed in order to lower the energy threshold,  down to an  effective  electron kinetic  energy  of $T_{\rm eff}  =  3.5 \,  {\rm MeV}$.  There was a significant improvement in the determination of the neutrino mixing parameters and in the extraction of the total $^8B$ neutrino flux, which was measured to be 
$\Phi_{NC}= (5.14^{+0.21}_{-0.20}) \times 10^6 cm^{-2} s^{-1}$.
The analysis of LETA plus SNO III gave a  best fit value compatible with the LOW region of parameter space,  but the significance level was very similar to the one of the usual LMA solution. A global fit, including all  the solar experiments and  the  data obtained by KamLAND,  essentially  confirmed, instead,  the previous results\cite{SNOIII} for $\Delta m_{12}^2$.  This analysis made possible a further  improvement  in  the  angle  determination, giving, in  a  3  flavor analysis, the following best fit values~\cite{SNOPHYSUN}\,: 
$\rm{tan}^2 \theta_{12}=0.468_{-0.033}^{+0.042}; \, \Delta m_{12}^2=\left(7.59 ^{+0.20}_{-0.21}\right) \times 10^{-5} eV^2;
 \, sin^2 \theta_{13}=\left(2.00^{+2.09}_{-1.63}\right)\times 10^{-2}.$ 

In the same years also the Super-Kamiokande Collaboration presented new analyses, which essentially confirmed
the previous results of the so called SK-I phase. The SK-I data, corresponding to 1496 live days until July 2001, were relative to Elastic Scattering (ES) above 5 MeV and their analysis showed that the day-night asymmetry was    
consistent with zero within $0.9 \, \sigma$ and in agreement with the predictions of the LMA solution and there were no signals of 
spectrum distortion and of anomalous periodic variations of the rate (apart from the expected seasonal variation due to Earth's 
orbit eccentricity). The following analyses included data  of the different experimental phases  know as Super-Kamiokande II \cite{SKII}\,(from December 2002 to October 2005) and Super-Kamiokande III (from   July   2006   to  August   2008)\cite{SKIII}\,. These two ph,ases  substantially confirmed  the  SK-I  results for what concerns the absence of significant spectral distortion, the total $^8B$ measured flux and the day-night asymmetry.

Since  September  2008,  Super-Kamiokande  is  running  with  a new  data acquisition system and electronics and this allowed a wider dynamic range in the measured charge. This phase of the experiment, denoted as  Super-Kamiokande-IV\cite{SuperKamiokandeIV}\,, is characterized by an optimization of the selection criteria and in particular by the introduction of a new event selection parameter,  the \lq\lq multiple scattering goodness\rq\rq,  which has the effect to reduce the very low energy background events (like the ones from $^{214} Bi$). Thanks to the adoption of this selection criteria in the analysis below $7.5 MeV$ of SK III and SK IV data it has been possible to reach a better signal to noise ratio. In addition to this, a small mistake in the $^8B$ energy spectrum shape (with impact on the flux) has been found by the SK Collaboration and fixed in this new analysis, which includes an updated live time of 1069 days. 
The kinetic energy  threshold has been slightly lowered, down to 3.5 $MeV$, and could be further reduced in the near future.   The total systematic error has been reduced to 
$1.7 \%$ (from the $3.2-3.5 \%$ of SK-I and the $2.1 \%$  of SKIII). 
The value measured for elastic scattering would correspond, in absence of oscillation, to a $^8B$ neutrino flux equal to 
$\left(2.34 \pm 0.03 (stat.) \pm 0.04 (syst.)\right)\times10^6 cm^{-2} s^{-1} $.  
The day-night asymmetry is still consistent with zero within $2.3 \, \sigma$.

Quite recently the SNO Collaboration performed a combined three phases analysis, in which all the data of the experiment were combined into a single dataset~\cite{SNO3fasi}\,. The study took advantage from the use of a pulse shape analysis (making possible a better particle identification) and a signal extraction based on a survival probability fit independent on any flux model. The flux determination extracted from the signal\cite{SNOPHYSUN}\,,$\Phi_{^8B}=\left(5.25 \pm 0.16^{+0.011}_{-0.013}\right)\times10^6 cm^{-2} s^{-1}$,
 is in good agreement with the SNO-LETA results. The total uncertainty is slightly larger than $3 \%$ and it is independent on the 
Solar Standard Model and on the oscillation models 
The oscillation parameter study is also improved with respect to previous analyses\footnote{For the exact determination of the mixing parameter values see~\cite{SNOPHYSUN}\,.}. The analysis is presently under revision, to include recent $\theta_{13}$ measurements and the updated results should be published soon. Once more, the SNO contribution is dominant for the $\theta_{12}$ determination, but it is not sufficient by itself to exclude the LOW solution, which is anyhow eliminated by Borexino and KamLAND independently.
In addition to the important role played in the determination of $^8B$, SNO also aims to improve the study of hep neutrinos, which represent the highest energy and lowest flux contrbution of the solar neutrino spectrum. 
 
%\subsection{Impact of  KamLAND experiment on solar neutrino physics}
The KamLAND experiment  played a fundamental role in solar neutrino physics, starting from its first data~\cite{firstKL}\,,  that were determinant to  prove the validity  of the oscillation hypothesis showing that the LMA solution was the correct one.
A new  set of  data were  collected between  March 2002 and  January 2004  with an important upgrade on  the detector, in which the photocatode  coverage was increased  and the energy resolution improved. 
Furthermore  an improved analysis, 
%technique in term of 
with a background reduction made possible by better techniques in the event selection cuts based on the time,  position  and  geometry  of  the  events, was performed. 
%A study, including data obtained in the 2002-2004 campaign and data obtained with a re-analysis of the previous data was performed\cite{KL-2004}. 
The study included the data obtained in the 2002-2004 campaign and a re-analysis of data obtained before. 
The observed events  above  $2.6  \,   {\rm  MeV}$  were 258, against an  expected number of antineutrino events in  absence  of antineutrino  disappearance  equal to  $365.2  \pm  23.7  {\rm(syst)}$; 
this corresponds to a $\bar{\nu}_e$  survival probability equal to $0.658 \pm 0.044 ({\rm stat}) \pm 0.047 ({\rm syst})$.
% Above  $2.6  \,   {\rm  MeV}$  the  expected number of antineutrino events in  absence  of antineutrino  disappearance  was 
% $365.2  \pm  23.7  {\rm(syst)}$  against  the 258 observed events; this corresponds to a $\bar{\nu}_e$  survival probability equal % to $0.658 \pm 0.044 ({\rm stat}) \pm 0.047 ({\rm syst})$.
The  best  fit  obtained from this analysis was  in the so-called  LMAI region with values  of $\Delta m_{12}^2$ around $8 \cdot 10^{-5} \, {\rm eV}^2$.
Including in the KamLAND analysis the results coming from  the solar neutrino experiments, the allowed  values of the angle were restricted  and the two flavor combined  analysis gave
 $\Delta m_{12}^2=7.9^{+0.6}_{-0.5}\times 10^{-5}   \,    \rm{eV}^2$ , \,$\rm{tan}^2  \theta_{12}   = 0.40^{+0.10}_{-0.07}$ at a  $1 \, \sigma$ level~\cite{KL-2004}\,.

The  following KamLAND analysis included new data  collected up  to May 2007~\cite{KLfollowing2}\,; it was obtained with a further reduction of  systematic  uncertainties and of  background. 
Furthermore the radius of the fiducial volume was increased from 5.5 to 6 m.
The  allowed oscillation parameter  values were  $\Delta  m_{12}^2=\left(7.58^{+0.14}_{-0.13}  (\rm{stat})  ^{+0.15}_{-0.15}
(\rm{syst})\right)\times10^{-5}  \, \rm{eV}^2$  and  
$ {\rm  tan}^2  \theta_{12}=0.56^{+0.10}_{-0.07}  ({\rm stat}) ^{+0.10}_{-0.06}  ({\rm syst})$,  
for ${\rm  tan}^2 \theta_{12} < 1$. 
In the three neutrino oscillation analysis the main effect  was to enlarge the uncertainty on  $\theta_{12}$, leaving $\Delta m_{12}^2$  substantially unchanged. 

 %\subsection{Borexino}\label{Borexino}\
In the last five years a new important step forward was made possible by  Borexino Collaboration, which  started the data taking in May~2007. Borexino, thank to the ability to reach a very low radiopurity level at the ton scale~\cite{14-ctf3}\,, is the first real time experiment investigating the low and medium energy part of solar neutrino spectrum. 

One of its main focus is the study of    
 the  ${^7}Be$ monochromatic line. In the first analysis performed by the Collaboration the  ${^7}Be$  signal was extracted 
from the background using data collected in 47.4~live  days.  The best value  estimate for  the rate  was $47  \pm 7  \, ({\rm stat})  \pm  12  \,  ({\rm syst})$~counts/(day~$\cdot$~100~ton),  where  the systematic   error  was  mainly   due  to   the  fiducial   mass  determination~\cite{bxfirstresults}\,. A second release of data was reported after 9 months from an  analysis of 192 live days, corresponding to 41.3~ton$\cdot$yr fiducial exposure to solar neutrinos.
The  total estimated  systematic error  was~\cite{Arpesella2008}  8.5\%. The best value for the  interaction rate of the 0.862~MeV $^7Be$ solar   neutrinos   was   $49  \,   \pm   3   ({\rm   stat})  \pm   4   ({\rm syst})$~counts/(day$\cdot$100~ton), in very good agreement with the predictions  of the LMA oscillation solution: $48 \, \pm 4$~counts/(day$\cdot$100~ton).

In order to reduce the systematic uncertainties a calibration campaign was performed in 2009  introducing  inside  the detector  several  internal  radioactive sources $\alpha$'s, $\beta$'s, $\gamma$'s, and  neutrons, at different energies and in hundreds of different  positions. 
Thanks to this calibration  campaign, the  systematic error  was reduced  to $2.7\%$  and the total uncertainty to $4.3\%$.
The data  set run from May 2007  to May 2010, with  a fiducial exposure equivalent  to 153.6 ton$\cdot$year. The $^7Be$ solar  neutrino rate was  evaluated to be  ${\rm 46.0 \pm 1.5  (stat) \pm 1.3(syst)}$~counts/(day$\cdot$100~ton)~\cite{Bellini2011rx}\,. 

Another important contribution of Borexino to the knowledge of solar neutrinos has been the measurement of the fluxes of {\it pep} and {\it CNO} neutrinos, that had never been directly measured before. These measurements are important because in the Solar Standard Models the flux of {\it pep} neutrinos is predicted with a small uncertainty (around $1 \%$), due to the solar luminosity constraint and to their direct connection to {\it pp} neutrinos; therefore its experimental determination would be a stringent test of the validity of these astrophysical models. 
On the other hand, the detection of neutrinos within the {\it CNO} bi-cycle is central to probe the solar core metallicity and contribute in this way to the solution of the solar metallicity problem~\cite{Serenelli:2011py, Basu}\,.
%\cite{SSM}.

The expected rate from \mbox{\it pep}  and \mbox{\it CNO} neutrino interaction is on the order of a few counts per day in a 100\,ton  target.  
In order to detect \mbox{\it pep}  and \mbox{\it CNO} neutrinos the Borexino  Collaboration adopted  a novel  analysis procedure to suppress the main source of background  in the  1-2  MeV  energy range,  which is due to  the cosmogenic $\beta^+$-emitter \mbox{$^{11}$C}  (lifetime of 29.4 min) produced within the scintillator  by muon  interactions with {\mbox{$^{12}$C}}  nuclei.  
The  background due to \mbox{$^{11}$C}  can be reduced by performing a space and time veto following coincidences between signals from  the muons  and  the cosmogenic  neutrons~\cite{pep-ctf}\,. This technique (Three-Fold Coincidence, TFC) is  based on the reconstructed  track  of  the  muon  and the  reconstructed  position of  the neutron-capture $\gamma$-ray. This criterium of rejection gave a rate of (2.5$\pm$0.3)  counts  per  day due to muons, corresponding to about (9$\pm$1)$\%$  of the  original  rate,  while preserving 48.5\% of the initial exposure.
Thanks to a small  difference in  the time distribution of  the scintillation signal  that arises from the finite  lifetime of ortho-positronium as well as  from the annihilation $\gamma$-rays, which present  a distributed, multi-site  event topology and a  larger average ionization  density  than electron interactions, it is possible to discriminate  \mbox{$^{11}$C} $\beta^+$ decays  from neutrino-induced $e^-$recoils  and $\beta^-$ decays  exploiting the pulse  shape differences  between  $e^-$ and $e^+$  interactions  in  organic liquid  scintillators.

The  Borexino Collaboration  presented the results in an analysis, published  in 2012~\cite{pepBX}\,,  based  on a binned likelihood multivariate fit performed
on  pulse  shape, spatial  distributions and energy.
In the  energy region of interest a fit  procedure was applied to radioactive backgrounds and to the contribution from \mbox{\it  pp} solar  neutrinos, that was  fixed to  the SSM  assuming  a MSW-LMA value of~\cite{pdg2010}\,$\tan^2\theta_{12}$=0.47$^{+0.05}_{-0.04}$,  $\Delta m^2_{12}$={(7.6$\pm$0.2)}$\times10^{-5}$\,eV$^2$,  and  to the contribution   from  $^{8}$B  neutrinos to  the  rate  from   the  measured flux of LETA and SNOI+II+III.
The  obtained results  for the  \mbox{\it pep}  and \mbox{\it CNO} neutrino interaction  rates, in units of  $\mbox{counts/(day$\cdot$100\,ton)}$, are $3.1  \pm 0.6_{\rm stat} \pm$ 0.3$_{\rm syst}$ and $<7.9$  ($<7.1_{\rm  stat\,only}$)  respectively, corresponding to a solar-$\nu$ flux of $\left(1.6\pm 0.3 \right)\times10^{8}  cm^{-2} s^{-1}$ for \mbox{\it pep} and $<7.7 \times 10^{8}  cm^{-2} s^{-1}$ for \mbox{\it CNO}.

\section{The mass and mixing pattern}
In many cases the solar neutrino data global analysis have been performed in the two flavor approximation. This simplifying assumption was justified  by the fact that the upper limit derived by the reactor experiments  for the 
mixing angle between the first and the third mass generation was quite severe and, therefore, the results of the analysis
done with $\theta_{13} = 0$ were presumably a good approximation. A recent example can be found, for instance, 
in ~\cite{SNO3fasi}\,. In that case, as already reported in section (\ref{last-decade}), the main results 
obtained by SNO collaboration  also in the two flavor analysis were the following: the SNO data
 are fundamental to determine the value of the $\theta_{12}$ mixing angle, but it is essential to include also the results 
of the previous solar neutrino experiments and of Borexino and the ones from KamLAND to select definitely the LMA solution 
and to restrict significantly the allowed values of $\Delta m_{12}^2$ . The small tension emerging  between the solar neutrino 
results and the KamLAND data (that would prefer slightly larger values for $\Delta m_{12}^2$) is reduced in the full 
three flavor analysis. The best fit is obtained for values of $\theta_{13}$ different from zero 
($sin^2 \theta_{13} = 0.025^{+0.018}_{-0.015})$ and the corresponding selected regions in the 12 mixing parameter plane 
are ~\cite{SNOPHYSUN}\, :
	$tan^2 \theta_{12} = 0.446^{+0.048}_{-0.036} \, \,  ; \,  \Delta m_{12}^2=\left(7.41^{+0.21}_{-0.19}\right)\times10^{-5} {\rm eV}^2 \, .$ 
 
The indication  in favor of values of $\theta_{13}$ different from zero was in agreement also 
with the outcome of recent theoretical analyses~\cite{Fogli:2011qn}\,and was strongly confirmed  during this year by a series of  results from the long baseline experiments T2K~\cite{T2K}\,and MINOS~\cite{MINOS}\,and,
above all, by three different  neutrino reactor experiments~\cite{DoubleCHOOZ, DayaBay,RENO}\,, which found values of $sin^2 \theta_{13}$ centered between 0.020 and 0.030.
 
 %\subsection{Status of the mixing parameters determination in a 3 flavor analysis}
\label{14-sec:status-mixing}

Both in two and three flavor analyses, the  higher values  of  $\Delta m_{12}^2$  in  the LMA  region were  excluded,
together with the full LOW  solution, mainly thanks to the large discrimination
power of KamLAND. This experiment, however, did not contribute significantly to
improve  the mixing  angle determination  and the  uncertainty on  this parameter
remained quite high.   For a comparison between the results obtained in the two and in the three  flavor  analyses one can 
look for instance at the Tables IX and X of \cite{SNO3fasi}\,, in which the results of the recent analysis of all the solar + KamLAND results, performed by the SNO collaboration, are summarized. 

As already said, the precise determination of  the $^8B$ solar neutrino flux, 
$\Phi_{^8B} =\left(5.25 \pm 0.16  (\rm stat)^{+0.011}_{-0.013}  (\rm  syst)\right) \times 10^6\,{\rm cm^{-2} \, s^{-1}}$,  made possible by the combined  analysis of  the different  SNO phases,  presents a significant reduction of the  systematic uncertainty.
This  result was consistent  with, but  more precise  than, both  the {\em high-Z}
BPS09(GS), $\Phi =  (5.88 \pm  0.65) \times 10^6  {\rm \, cm^{-2} \, s^{-1}}$,  and 
{\em low-Z} BPS09(AGSS09),
$\Phi  =  (4.85  \pm  0.58)  \times  10^6  {\rm \, cm^{-2}  \,  s^{-1}}$,  solar model predictions \cite{14-serenelli2009yc}\,.

Recently different groups published the results of global phenomenological studies of the mass and mixing parameters
(see~\cite{14-Fogli:2012ua, 14-Tortola:2012te,GonzalezGarcia:2012sz,Smirnov:2012ei})\,, finding a substantial agreement. We refer the interested  reader to these papers for a detailed analysis of mixing parameters and  
for  a discussion about the promising perspectives for leptonic CP violation searches opened by the interval of allowed values for $sin^2 \theta_{13}$ selected by  the recent experimental results~\cite{14-Fogli:2012ua}\,.

%and the  possible consequences of  the recent experimental results  obtained for 
%$sin^2 \theta_{13}$ and the  promising perspectives 
%that the interval of allowed values for this parameter (quite significantly different from zero) opens for future  experiments 
%looking for  leptonic CP  violation \cite{14-Fogli:2012ua}.

The combination of  the recent SNO collaboration's analyses \cite{LETA, SNO3fasi} and
of the Borexino measurements \cite{14-Bellini:2008mr} made possible a detailed
study of  the low energy  part of the  $^8B$ solar neutrino spectrum.  Even if
characterized  by a larger uncertainty (mainly due  to a more limited
statistics), Borexino data confirm  the LETA indication of low energy data
points  lower  than  the  theoretical  expectations  based  on  matter  enhanced
oscillation and solar models.     
These   results   were in agreement    also   with   the Super-Kamiokande observation  
%\cite{14-SKI-2005} 
of a flat  spectrum,  consistent with the undistorted spectrum hypothesis.  
The emergence of this slight tension
between theory and experiments seems to indicate  the presence of new
subdominant  effects  and  also suggests   the possibility  of  non-standard
neutrino interactions (like those studied in \cite{Friedland:2004pp}) or
the  mixing  with a  very  light  sterile neutrino \cite{deHolanda:2010am, Smirnov:2012ei}\,.
Future solar  neutrino experiments, like SNO+,  could shed more  light on this
subject, by performing precision measurements of lower energies solar neutrinos 
(like the \mbox{\it pep} neutrinos).

The accuracy reached by the data obtained at different solar neutrino experiments suggests also the possibility of using 
these results to test in an independent way the neutrino propagation and mixing models from one side and the astrophysical 
models ruling neutrino production on the other side. With this aim it is possible to perform a so called \lq\lq free flux analysis\rq\rq, in which one lets the different solar neutrino fluxes vary in order to accommodate the experimental data, maintaining the functional 
dependences as predicted by the standard models and assuming a \lq\lq luminosity constraint\rq\rq (which assures that the fusion processes
are responsible for solar luminosity and guarantees the conservation of energy for nuclear fusion of light elements).
The main aspects and the results of such an analysis are discussed for instance in\cite{noi}\,.
The precision of the $^7Be$ and $^8B$ neutrino fluxes is driven by the Borexino and SNO (SK) neutrino experiments, while the precision of
 the {\it pp} and {\it pep} neutrino fluxes at present mainly comes by the imposition of the luminosity constraint. The neutrino data directly
demonstrates that the Sun shines by the {\it pp}-chain and the {\it CNO} bi-cycle only contributes to the total luminosity at the percent level
(around $0.8\%$ and $0.4\%$, respectively for {\em high-Z} and {\em low-Z} solar models).
If one further relaxes the assumption, giving up the luminosity constraint, one gets an estimate of solar luminosity inferred by neutrino data 
which is in agreement with the directly measured one within about $15\%$.

\section{Future of  solar neutrino physics}

In the near future the study of solar neutrinos, for what concerns the pp-chain,  will be focused on the low energy part, which  represents the great majority of 
the spectrum, but up to now  has been an almost unexplored realm.
A significant contribution is expected from  Borexino and SNO$+$  \cite{14-Kraus2010zzb}  experiments.  
The Borexino Collaboration already  proved  its capability to perform the first measurements of {\it pep} and  {\it CNO} neutrinos. Since July 2010 a big effort was undertaken, with a new purification campaign, in order to further reduce the  main radioactive background sources. 
Being SNOLAB located twice deeper underground than the Gran Sasso laboratory, the SNO+ experiment, that should start taking data soon, will take advantage from a lower  muon flux and hence a strongly reduced $^{11}$C rate.  This  could determine a fundamental  improvement  in the {\it  pep} neutrino  measurement, where  a  $5  \%$ uncertainty is expected.
Both collaborations hope to  attack the main problem of measuring the lowest energy  parts of the solar neutrino spectrum, the {\it pp}  neutrinos and the $0.38 \, {\rm MeV}$  Beryillium line, even if  the  presence of $^{14}$C background  in  the organic  scintillators  will make this very  low energy measurements a very hard task.

The pile up effects of $^{14}$C are being studied with Monte Carlo simulations and using the data themselves. The Borexino Collaboration is planning to design also a possible test of  these pile-up events by inserting $CO_2$, in gaseous form, in the scintillator. This would increase of a factor of about 2 the contribution from $^{14}$C and then the pile up induced from it. In this way, comparing the data before and after the $CO_2$ insertion, it could be easier to disentangle the $^{14}$C pile up effects.

All future experiments aiming to  measure the low energy part of  solar neutrino spectrum are characterized by  a very large  detector target mass and  by the need to reach  very high radiopurity levels, in order to be suitable for the detection of a low rate signal in a region characterized by different potential radioactive background sources. This requirement is common also to the experiments looking for neutrinoless double $\beta$ decay or for dark matter signals (search for signatures
of WIMPs) and, therefore, many of the planned solar neutrino experiments are multipurpose experiments
designed also for the other above-quoted topics.

Thanks to  the  experience acquired with Borexino, a new generation of neutrino detectors has  been proposed.  LENA (Low  Energy Neutrino  Astronomy) \cite{14-LENA} is a multipurpose detector  aiming to study, among others issues, solar neutrinos.  The project consists in a cylindrical detector with a diameter of 30~m and a
length of about 100~m for about 50 kilotons of liquid scintillator as target mass. The light is collected  by about  45,000,  20 cm in diameter, photomultipliers equipped with conic mirrors. The corresponding surface coverage is about 30\% and the solvent for the liquid scintillator will probably be linear alkylbenzene which has a high  light yield and large attenuation length  on the order of  10 to 20~m at a wavelength of 430~nm. The photoelectron yield  is about 200 photoelectrons per MeV for a scintillator mixture containing 2g/l PPO  and 20 mg/l  bisMSB as wavelength shifters. As  alternative solvent  option LENA could employ the PXE  \cite{14-PXE}  or a mixture of PXE and dodecane.
A high statistics can be reached in short times and in both Pyhsalmi and Frejus underground laboratories where the muon flux is lower compared to LNGS underground laboratories. For  {\it pep},  {\it CNO}  and low-energy $^{8}$B-$\nu$s detection a fiducial mass  of $\sim$30\,kton is mandatory, while the fiducial mass for $^{7}$Be-$\nu$s and  high-energy ($E>5$\,MeV) $^{8}$B-$\nu$s could be enlarged to 35\,kton or more. The expected rates (for the channel $\nu e\to e\nu$) in 30\,kton for {\it pp} neutrinos,  {\it pep} neutrinos and the {\it CNO} bi-cycle, using the most recent  solar model  predictions are 40 cpd (counts per day), 280 cpd and 190 cpd respectively; while for $^{7}$Be-$\nu$s and $^{8}$B-$\nu$s they are 100 cpd and 79 cpd in 35\,kton fiducial mass \cite{14-LENA}\,.

The Low Energy Neutrino Spectroscopy (LENS) detector has as main goal the real time measurement, as a function of their energy, of solar neutrinos and 
particularly of the {$\it  pp$} ones. 
To make an energy spectrum measurement on low energy neutrinos, one has to reach a low threshold for the charged current (CC) process and discriminate the  radioactive background.  The CC process employed in LENS is the neutrino induced  transition of $^{115}$In to an excited state of $^{115}$Sn (i.e. $\nu_{e} + ^{115}{\rm In} \rightarrow ^{115}{\rm Sn}^* + e^{-}  \ ({\rm E=E_\nu -114 keV})$) followed by $^{115}{\rm Sn}^*  (\tau = 4.76 \mu s) \rightarrow ^{115}{\rm Sn + \gamma(498 keV) + \gamma(116 keV)}$.
Thanks to this reaction it is possible  to detect low energy neutrinos with a threshold of 114 keV and measure their energy. The primary interaction and the secondary cascade give a triple coincidence, correlated  in time and  space. The detection medium is a liquid scintillator chemically doped with natural indium ($^{115}$In  = 95.7\%). LENS  should be able to  determine the  low energy  solar neutrino fluxes with an accuracy $\leq 4 \%$ \cite{14-Raghavan2004}\,.
   
Another interesting possibility, under investigation by different collaborations, is that of using  scintillation detectors with liquid noble  gases,  like xenon,  argon  and  neon.  These materials are relatively inexpensive, easy to obtain
and dense; moreover, they can be  quite easily purified, have a high scintillation yield (about  $30-40$ photons/keV) and do not absorb their own scintillation light.

A first example is given by the CLEAN/DEAP  family,  a  kind of detectors  based entirely on scintillation  in liquid  neon (LNe)  and liquid argon (LAr). Some prototypes of this kind have been  installed in the SNOLAB.
The CLEAN (Cryogenic Low Energy Astrophysics with Noble gases) detector \cite{14-CLEAN2004} will  be made by  a stainless steel  tank, of about 6  meters of diameter,  filled with 135~metric tons of cryogenic  liquid neon; only  the  central  part  of  it,  surrounded isotropically  by  a  series  of photomultipliers, will  constitute the detector fiducial  volume.  An external water tank will  act  as $\gamma$-ray and  neutron shielding  and muon  veto.
A statistical uncertainty on  the ${\it pp}$ measurements of the  order of $1 \,\%$ is foreseen.

The  XMASS experiment  \cite{14-Moriyama2011zz} is a  multipurpose experiment, mainly focused on neutrinoless double $\beta$ decay and dark matter searches, but designed also to  study $\it pp$ and $^7Be$ neutrinos.
It will employ liquid xenon and should reach very low levels of background  and energy threshold.
The full  XMASS detector will  have a fiducial volume of  10 metric tons. 

Another  experimental  project  based on  the  noble gases  liquid scintillator technique is  that of DARWIN (DARk matter  WImp search with Noble
liquids)  \cite{14-DARWIN}\,, conceived for the study for a future multi-ton scale  LAr and LXe dark  matter search facility in  Europe. 
The main goal of the experiment  is to look for a WIMP signal, but the energy region of the nuclear recoil  spectrum, below 200~keV, that should be
investigated by this future experiment  is of particular interest also for the study of the ${\it  pp}$  solar neutrinos.

\section{Conclusions}

%\section{\textcolor{red}{Conclusion OSSIA 14.8+14.9 di REWantonelli - Da Fare tutto il capitolo}}

In the last 10-15 years very important steps forward have been done in the study of solar neutrinos and consequently 
in our knowledge of neutrino properties. The results obtained around the turn of the century by the \v{C}erenkov experiments SuperKamiokande and SNO and the confirmation offered by the KamLAND experiment monitoring the reactor antineutrinos  were determinant to solve the long standing Solar Neutrino Puzzle, proving in a crystal clear way the validity of the oscillation hypothesis 
and selecting the Large Mixing Angle solution in the mixing parameter space. In the following years the aforementioned 
experiments went on producing data improving their statistics, reducing the systematic errors with the introduction of new detection techniques (for instance in the case of the neutral current studies performed by the SNO collaboration) and improved 
methods of statistical analysis and in some cases lowering their energy threshold. 
In the last five years a new piece of the puzzle was added, thanks to Borexino experiment, which performed the first real time
measurement of the low energy part of solar neutrino spectrum, observing with increasing accuracy the monochromatic beryllium 
line and obtaining the first measurements of {\it pep} and {\it CNO} neutrinos.
Meanwhile a long standing question found its answer during the last year with the discovery at different long baseline reactor experiments that the $\theta_{13}$  mixing angle is definitely different from zero.

A generally coherent picture emerges from  all of these experimental achievements and from the global analyses 
performed by the single experimental collaborations and in other theoretical and phenomenological studies and  
the basic properties and the values of  the parameters driving solar neutrino oscillations, $\Delta m^2_{12}$ and $\theta_{12}$, 
are now known with quite a satisfactory accuracy.  
%(Tables~\ref{tab:global2nu}~and~\ref{tab:chi3nu}.
Nevertheless, there are still important aspects of the oscillation mechanism on which it would be desirable to shed more light. In particular the transition between the low energy region, where the oscillation takes place essentially in vacuum, and the higher 
energy regime,  in which matter interactions become extremely relevant, requires more study.  
In fact the data available up to now for the low energetic $^8B$ neutrinos (which are affected by a significant uncertainty) don't show explicitly the rise of the Solar neutrino spectrum\cite{Chavarria} that,  according to the LMA solution, should appear when one decreases the energy passing from the matter dominated towards the vacuum region. More data coming from Super-Kamiokande, Borexino 
and SNO+ experiments are expected to further explore the conversion in this regime.
The precise measurement of low energy neutrinos like {\it pep} (and the comparison with eventual future {\it pp} neutrino measurement), exploiting the fact that {\it pep} neutrinos are more energetic than  the $^7Be$ ones, could also help to see small solar matter effects in the flavor conversion. The {\it CNO}  neutrinos would be even more suitable for this kind of studies, because their energy is of the same order of the {\it pep} ones, but they are produced at higher temperatures and therefore at higher densities, leading to even more significant matter effects.

The combination of all of these studies and of the results  by reactor experiments is expected to eliminate in the near future  
this uncertainty, clarifying if the oscillation mechanism is well understood or one has to invoke the so called non standard 
interactions.
 
%From an astrophysical point of view, the accuracy reached by the solar neutrino experiments is, at least, partially sufficient 
%to realize the original Davis and Bahcall's dream and use neutrinos to probe the mechanisms driving the fusion processes in the 
%Sun. 
From an astrophysical point of view, the accuracy reached by the solar neutrino experiments is for many aspects sufficient 
to realize the original Davis and Bahcall's dream and use neutrinos as probes to get a better insight into the mechanisms driving 
the fusion processes in the Sun. 
 Generally speaking the Standard Solar Model predictions are in excellent agreement with solar fluxes, regardless of the solar composition assumed in the construction of the model.  
However the measured values of the fluxes for the different components of the {\it pp}-fusion chain unfortunately fall somehow in the middle\cite{Serenelli:2011py}  between the predicted values for the two different versions of the solar models usually denoted as  {\em high-Z} and {\em low-Z} solar models and, therefore, it is very unlikely that the study of neutrino fluxes from the {\it pp}-chain alone will be able to discriminate between these different solar compositions. 
An important possibility could be offered by the combined analysis of the different components of pp-chain and of {\it CNO} 
neutrinos (which can bring important information about the solar core). Borexino recently opened a new way, establishing the most stringent upper limit on 
{\it CNO} fluxes which is about a factor of 1.5 larger than solar model predictions and the recent developments, also in background  subtraction techniques, offer to this experiment the possibility of a further improvement. 
In this regard, the potentialities of SNO+ are even more promising, thanks to its location deeper underground and to the higher detector mass, but at present the primary solar neutrino measurements are planned to take place after the double beta decay measurements are carried out for at least 4 years. 

This issue is relevant because it could make possible an important step forward in the solution of the so called solar metallicity 
problem\cite{Serenelli:2011py, Basu}\,. In the last decades three-dimensional radiation hydrodynamic (3D RHD) models of the solar atmosphere have been consistently developed by different groups\cite{3DRHD} \cite{3DRHD2} \cite{3DRHD3}. 
Their main impact on solar models and solar neutrino physics is that 
the composition of different elements have been significantly reduced (down to $30-40 \%$  for some key elements, like C, N and  
O). However these models, usually denoted as {\em low-Z models}, seem to be definitely less efficient than  previous
 {\em high-Z} models in matching helioseismic constraint, like the internal sound speed and density profiles and the depth of the 
solar convective envelope. 
The different attempts developed in these years to solve this solar abundance, or solar metallicity, problem by modifying some
inputs of the Solar Standard Models partially failed, in the sense that  a simultaneous  solution to all the problems of consistency 
with helioseismology has not yet been found. 
A possible way out could be offered by an increase of radiative opacities, but the increase of this factor needed to restore 
the agreement with helioseismology seems be much higher than the estimated present uncertainty of its present calculations 
in the {\it low-Z} models.
It's worth noticing that  the $^7Be$ and $^8B$ fluxes of {\em  low-Z} SSMs  with increased  opacity would essentially coincide 
with  those from  a {\em  high-Z} model \cite{14-serenelli:2010}\,,  showing the intrinsic degeneracy between composition and opacities.
By  using   $^8B$  and   now  $^7Be$  as   thermometers  of  the   solar  core
\cite{14-haxton:2008}\,,\,{\it CNO} neutrinos  represent a  unique way  to  
break this degeneracy and provide an independent  determination of the 
{\it CNO} abundances,
particularly the C+N  abundance in  the solar core.  Keeping in mind  the 
antagonism between solar interior  and solar atmosphere models that  the solar abundance
problem  has established,  results from {\it CNO} fluxes  will be  of  the 
outmost relevance for solar, and by extension stellar, physics. 

 A precise measurement of {\it CNO} neutrinos would have a great impact on our knowledge of the formation and evolution mechanism
 not only of the Sun, but also of other stars and of the planets in the solar system.
 First of all the detection of these neutrinos would confirm directly that the 
{\it CNO} bi-cycle is an important source of energy for the 
stars (which can become the dominant fusion mode for stars with masses right above the solar one). Moreover, an accurate determination of the combined $^{13}N+^{15}O$ flux and consequently of the solar {\it C+N} abundance 
would be extremely useful not only for the solution of the solar abundance problem, but also for a better knowledge of the 
mixing mechanism involved in the evolution of the Sun and in the earlier phases of planet formation. For a more detailed discussion of this topic we refer the interested reader to \cite{noi}\,.
 
\section*{Acknowledgements}
It's a pleasure for us to thank A. M. Serenelli and C. Pe\~na Garay, who realized with us the paper \cite{noi}\,, on which this review is partially based.
%\section{References}
 % Versione bibliografia nuova

\end{document}